\documentclass[preprintnumbers,amsmath,11pt,amssymb,floatfix,superscriptaddress,nofootinbib]{article}

\def\mytitle#1{\setcounter{equation}{0}
\setcounter{footnote}{0}
\begin{flushleft}\Large\textbf{#1}\end{flushleft}
\vspace{0.20cm}}
\def\myname#1{\leftline{{\large #1}}\vspace{-0.13cm}}
\def\myplace#1#2{\small\begin{flushleft}\textit{#1}\\
\texttt{#2}\end{flushleft}}

\def\myclassification#1{\small\noindent
Keywords : Dark Energy, Scale Factor, Redshift parametrization.
       #1\vspace{0.5cm}}
\usepackage{graphicx}
\usepackage{amsmath}
\usepackage{bigints}
\usepackage{amssymb}

\begin{document}
\mytitle{Barboza-Alcaniz Equation of State Parametrization : Constraining the Parameters in Different Gravity Theories}

\myname{$Promila ~Biswas$\footnote{promilabiswas8@gmail.com} and  $Ritabrata~
Biswas$\footnote{biswas.ritabrata@gmail.com}}
\myplace{Department of Mathematics, The University of Burdwan, Golapbag Academic Complex, City : Burdwan-713104, District. : Purba Burdwan, State : West Bengal, Country : India.} {}
 
\begin{abstract}
To justify the twenty years old distant Ia Supernova observations which revealed to us that our universe is experiencing a late time cosmic acceleration, propositions of existence of exotic fluids inside our universe are made. These fluids are assumed to occupy homogeneously the whole space of the universe and to exert negative pressure from inside such that the late time accelerated expansion is caused. Among the different suggested models of such exotic matters/ energy popularly coined as dark matter / dark energy, a well known and popular process is ``introduction of redshift parametrization'' of the equation of state parameter of these fluids. We, very particularly, take the parametrization proposed by Barboza and Alcaniz along with the cosmological constant. We use thirty nine data points for Hubble's parameter calculated for different redshifts and try to constrain the dark energy equation of state parameters for Barboza Alcaniz modelling. We then constrain the dark energy parametrization parameters in the background of Einstein's general relativity, loop quantum gravity and Horava Lifshitz gravity one after another. We find the 1$\sigma$, 2$\sigma$ and 3$\sigma$ confidence contours for all these cases and compare them with each other. We try to speculate which gravity is constraining the parameters most and which one is letting the parameters to stay within a larger domain. We tally our results of 557 points Union2 Sample and again compare them for different gravity theories. 
\end{abstract}

\myclassification{}

\section{Introduction}
Almost twenty years have had passed since the observational evidences from the distant SNeIa observation which concluded the late time cosmic acceleration of our universe, were collected \cite{cosmic_acceleration_paper_1, cosmic_acceleration_paper_2}. Understanding the constraints on the universe's expansion rate can be best studied by the study of the Hubble's parameter as a function of redshift, $H(z)$, which is a determining factor for the scale factor $a(t)$  introduced in the Friedmann-Lemaitre-Robertson-Walker(FLRW) metric. Hubble's parameter relates the Doppler shift measured velocity of several distant galaxies received from our planet with the distance of the corresponding galaxies which are located upto a few hundred megaparsec away from earth \cite{Riess_Observational Evidence}. Although, popularly attributed to Edwin Hubble, the famous Hubble-Lemaitre law ($V = H_0D$) was firstly derived from the Einstein field equations by Alexander Friedmann in 1922. Hubble actually gave a approximately corrected value to the constant $H_0$. Detailed studies of the scale factor via the studies of $H(z)$ allow us to probe the properties and to understand the natures of the fundamental components of the universe.

Upto a short distance from our solar system, we use Cepheid variables as standard candles. Type Ia supernova explosions (SNeIa) act as standard candles when we look towards distant galaxies \cite{cosmic_acceleration_paper_1, cosmic_acceleration_paper_2}. Besides this kind of utilisations of standard candles, the Baryonic Acoustic Oscillations (BAO) is used as standard rulers \cite{BAO1, BAO2}. With these two standard tools, study of the Cosmic Microwave Background(CMB) \cite{CMB1} has enriched the literatures produced in last twenty years, especially the developments of the standard $\Lambda$CDM cosmological model. However, these methodologies do not directly constrain the Hubble's parameter.

Besides these methods, another method, named as {\it ``Cosmic Chronometers"} method is suggested  by the references \cite{Jimenez_Loeb, cosmic_chronometers}. According to this method, the relative ages of old and passive galaxies, expressed as $~\frac{dz}{dt}$, can be used while we need to constrain the expansion history of the universe directly.

A sample of $\sim$ 11000 massive and passive galaxies has been analyzed and eight measurements of the Hubble parameter have been speculated with an accuracy of 5{\bf-}12\% in the redshift range 0.15$ < z < $1.1 in the reference \cite{Moresco}. Most {\bf of the} accurate constraints were found for low redshifts ($z <$0.3). Comparative studies of cosmic chronometers method with standard probes like SNeIa and BAO are found in the references \cite{Moresco2,Zhao,Wang,Riemer-Sorensen}. Some $H(z)$ points in the redshift range 0.35$< z <$0.5 are given in the reference \cite{Moresco_JCAP}.

We will give all these data in a tabular form in Table-I.

\begin{center}
~~~~~~~~~~~~~~~~~~~~~~~~~~~~~~~~~~~~~~~~~~~~~~~~~~~~~~~Table-I~~~~~~~~~~~~~~~~~~~~~~~~~~~~~~~~~~~~~~~~~~
\end{center}
\begin{minipage}{.5\linewidth}
\centering
\begin{tabular}{|| c | c  c  c  c ||} 
\hline
Sl No. & z & $H(z)$ & $\sigma(z)$ & Ref. No. \\ 
\hline
1 & 0 & 67.77 & 1.30 & \cite{Macaulay} \\
\hline
2 & 0.07 & 69 & 19.6 & \cite{Four_new_observational} \\ 
\hline
3 & 0.09 & 69 & 12 & \cite{Simon}  \\
\hline
4 & 0.1 & 69 & 12 & \cite{Stern}  \\ 
\hline
5 & 0.12 & 68.6 & 26.2 & \cite{Four_new_observational} \\
\hline
6 & 0.17 & 83 & 8 & \cite{Stern} \\
\hline
7 & 0.179 & 75 & 4 & \cite{Moresco} \\ 
\hline
8 & 0.1993 & 75 & 5 &  \cite{Moresco, cosmic_chronometers} \\
\hline
9 & 0.2 & 72.9 & 29.6 & \cite{Four_new_observational} \\
\hline
10 & 0.27 & 77 & 14 & \cite{Stern} \\
\hline
11 & 0.28 & 88.8 & 36.6 & \cite{Four_new_observational} \\
\hline
12 & 0.35 & 82.7 & 8.4 & \cite{Modeling_the_anisotropic} \\
\hline
13 & 0.352 & 83 & 14 & \cite{Moresco} \\
\hline
14 & 0.3802 & 83 & 13.5 & \cite{Moresco_JCAP} \\
\hline
15 & 0.4 & 95 & 17 & \cite{Simon} \\
\hline
16 & 0.4004 & 77 & 10.2 & \cite{Moresco_JCAP}  \\
\hline
17 & 0.4247 & 87.1 & 11.2 & \cite{Moresco_JCAP} \\
\hline
18 & 0.44 & 82.6 & 7.8 & \cite{WiggleZ_Dark_Energy} \\
\hline
19 & 0.44497 & 92.8 & 12.9 & \cite{Moresco_JCAP} \\
\hline
20 & 0.47 & 89 & 49.6 & \cite{cosmic_chronometers,Ratsimbazafy}\\
\hline 
\end{tabular}
\end{minipage}
\begin{minipage}{.5\linewidth}
\centering
\begin{tabular}{|| c | c  c  c  c ||} 
\hline
Sl No. & z & $H(z)$ & $\sigma(z)$ & Ref. No. \\ [0.5ex] 
\hline
21 & 0.4783 & 80.9 & 9 & \cite{Moresco_JCAP}\\
\hline
22 & 0.48 & 97 & 60 & \cite{Stern} \\
\hline
23 & 0.57 & 96.8 & 3.4 & \cite{Anderson}  \\
\hline
24 & 0.593 & 104 & 13  & \cite{Moresco}\\
\hline 
25 & 0.6 & 87.9 & 6.1 & \cite{WiggleZ_Dark_Energy}\\
\hline
26 & 0.68 & 92 & 8 & \cite{Moresco}\\
\hline 
27 & 0.73 & 97.3 & 7 & \cite{WiggleZ_Dark_Energy}\\
\hline
28 & 0.781 & 105 & 12 & \cite{Moresco}\\
\hline
29 & 0.875 & 125 & 17 &\cite{Moresco}\\
\hline
30 & 0.88 &  90 & 40 & \cite{Stern}\\
\hline
31 & 0.9 & 117 & 23 & \cite{Stern}\\
\hline
32 & 1.037 & 154 & 20 & \cite{Moresco} \\
\hline
33 & 1.3 & 168 & 17 & \cite{Stern}\\
\hline
34 & 1.363 & 160 & 33.6 & \cite{Raising_the_bar} \\
\hline
35 & 1.43 & 177 & 18 & \cite{Stern}\\
\hline
36 & 1.53 & 140 & 14 & \cite{Stern}\\
\hline
37 & 1.75 & 202 & 40 & \cite{Stern}\\
\hline
38 & 1.965 & 186.5 & 50.4 & \cite{Raising_the_bar}\\
\hline
39 & 2.34 & 222 & 7 & \cite{Delubac}\\
\hline
\end{tabular}
\end{minipage}

The observations of accelerated expansion are explained by introducing a new hypothetical energy component with negative pressure. This kind of exotic matter's name was popularly coined as Dark Energy (DE hereafter) or quintessence, which are generally characterized by the equation of state parameter (EoS), $\omega = \frac{p}{\rho}~~(< ~0)$, which is the ratio between the DE's pressure to its energy density (some good reviews can be found in the references \cite{Peebles, Padmanabhan, Copeland, Alcaniz, Biswas_P1}). The simplest and most natural possibility of the energy density among many proposed DE is satisfied as the quantum vacuum or introduced in EFE as the cosmological constant $(\Lambda)$ model. This particular interpretation of the cosmological term focuses on the unsettled situation in the particle physics/cosmology interface, in which the cosmological upper bound $(\rho_a \lesssim 10^{-47} GeV^4)$ differs from theoretical expectations $(\rho_a \lesssim 10^{71} GeV^4)$ by more than 100 orders of magnitude \cite{Weinberg, Sahni}. Thus, inspite of the fact that $\Lambda$ may be able to explain the majority of observations available so far, if DE is really associated with the vacuum energy density, we should search for a better explanation for the enormous discrepancy between observation and theory. Several tries to set an explanation have failed to become perfectly reasonable. These lead us, contrary to the beauty and simplicity of $\Lambda$, to introduce other proposals. Some of such models are time-varying cosmological term model \cite{ozer}, irreversible process of cosmological matter creation \cite{Lima}, Chaplygin gas family \cite{Kamenshchik, Biswas_P2} etc.

To describe DE EoS, some time dependent parametrizations are also proposed. These redshift dependent parametrizations can not be obtained from the scalar field dynamics as they are not limited functions, i.e., their EoS parameters do not lie in the interval defined by $\omega = \frac{\frac{\dot{\phi}^2}{2}-V(\phi)}{\frac{\dot{\phi}^2}{2}+V(\phi)}$, where $V(\phi)$ is the field potential. Nevertheless, as DE dominance is a recent phenomenon, this particular aspect in the table is important because it may be possible to obtain a quintessence like behaviour as a particular approximation when $z$ is not too larger. Two prior families of redshift parametrizations are
$\omega(z) = \omega_0 + \omega_1 \left(\frac{z}{1+z} \right)^n$
and
$\omega(z) = \omega_0 + \omega_1 \frac{z}{(1+z)^n}$, ${\bf n \in \mathbb{N}}$

In the reference \cite{Barboza_Jr.}, authors have proposed a new parametrization which does not belong to the above family of parametrizations given as
\begin{equation}\label{new_parametrization}
\omega(z) = \omega_0 + \omega_1 \frac{z(1+z)}{1+z^2}~~~~~~~~~~,
\end{equation}

The EoS is where $\omega_0$ is the EoS at present time (the subscript and superscript zero denotes the present value of a quantity) and $\omega_1 = \frac{d\omega}{dz} \Bigr\lvert_{z=0}$ gives a measure of how much time dependent the DE EoS is. So we are more interested to check our desired results using Barboza-Alcaniz parametrization. This parametrization is found to be well-behaved and bounded function of redshift throughout the entire cosmic evolution. This parameter model always helps us to study the distant future of universe at $z= -1$ as well as to the last scattering surface of the CMB. Another important feature of this model can be observed through the definition of deceleration parameter $q$.

We know that the deceleration parameter $q$ is related to the Hubble's parameter by,
\begin{equation}\label{deceleration_parameter}
q = -\frac{\dot{H}}{H^2}-1 = \frac{3}{2}(1+\frac{p}{\rho})-1 = \frac{1}{2}(1+3\omega)~~~~~~,
\end{equation}
where $\omega= \frac{p}{\rho}$. A positive value of the deceleration parameter, $q$ indicates deceleration whereas $q<0$ implies acceleration. Replacing $\omega(z)$ into equation (\ref{deceleration_parameter}), i.e, we can write the deceleration parameter $q(z)$ with the help of Barboza-Alcaniz parametric form \cite{Mamon} as :
\begin{equation}
q(z) = q_0 + q_1~ \frac{z(1+z)}{1+z^2}~~~~~~~,
\end{equation}
where $q_0$ is the present value of $q(z)$, $q_0 = \frac{1}{2}(1+3\omega_0)$ and the rest part is the variations of the deceleration parameter with respect to $z$, $q_1 = \frac{3}{2}\omega_1$. The authors of the reference \cite{Subhojit_Saha} have performed statistical analysis using observational data from SNeIa, BAO and CMB shift parameter and have estimated the values of $H(z)$ (from the age of high-z galaxies) and obtained the best fit values of the parameters to be $\omega_0 = -1.11$ and $\omega_1 = 0.43$ at $1\sigma$ confidence level .

We know that for $z\gg 1$, i.e., at high redshift, $q(z)$ reduces to $q(z)= q_0 + q_1$. The universe will be radiation dominated ($\omega = \frac{1}{3}$) by suitably choosing the values of $q_0$ and $q_1$ for $z\gg 1$ limit. For smaller values of $z$, it shows DE behaviours. We also know that for $z\ll 1$, this parametrization leads to linear parametrization $q(z)= q_0 + q_1 z$ \cite{A_G_Riess}. The main advantage of this model is that we can obtain finite values of $q$ in the entire range $z \in [-1, \infty) $ and is valid for entire evolution history of universe. One can use this parametrization for further studies of future evolution of universe also. This parametrization represents a good fit for low redshifts, but presents complicated properties for high redshift. It fails to explain the estimated ages of high redshift objects. We can open up some possibilities for future works regarding the nature of DE from this non-trivial parametrization.

For this new parametrization, the authors of the reference \cite{Barboza_Jr.} have deduced the bounds in $\omega_0 - \omega_1$ plane as :

For quintessence : $-1 \leq \omega_0 - 0.21 \omega_1$   and  $\omega_0 + 1.21 \omega_1 \leq 1$ (if $\omega_1 > 0 $)

$~~~~~~~~~~~~~~~~~~~~~~~~~~~~$and

$~~~~~~~~~~~~~~~~~~~~~~~~~~-1 \leq \omega_0 + 1.21 \omega_1$   and   $\omega_0 - 0.2 \omega_1 \leq 1$ (if $\omega_1 < 0$)

For phantom : $\omega_1 < \frac{-(1 + \omega_0)}{1.21}$ (if $ \omega_1 > 0$ )

$~~~~~~~~~~~~~~~~~~~~~~~~~~~~~$and

$~~~~~~~~~~~~~~~~~~~~~~~~~~~~~\omega_1 > \frac{(1 + \omega_0)}{0.21}$ (if $\omega_1 < 0$).

Quintessence, phantom, decelerated phase and some forbidden regions are efficiently classified in the $\omega_0 - \omega_1$ plane.

From the mass conservation equation, we have
\begin{equation}
\dot{\rho} + 3H(\rho + p) = 0 .
\end{equation}
For radiation ($p = \frac{1}{3}\rho$) the relation between density and redshift becomes :
\begin{equation}
\rho_{rad} = \rho_{rad,0}(1 + z)^4 .
\end{equation}
The matter ($p = 0$) density, as a function of redshift, turns to be
\begin{equation}
\rho_{DM} = \rho_{DM,0}(1 + z)^3 ,
\end{equation}
where $\rho_{rad,0} = \rho_{rad}(z = z_0)$, $\rho_{DM,0} = \rho_{DM}(z = z_0)$.
Using the equation of state given in equation (\ref{new_parametrization}), we get the DE density as a function of redshift given as,
\begin{equation}
\rho_{DE} = \rho_{DE,0} \bigg\{\frac{1+z^2}{(1+z)^2}\bigg\}^{\frac{3\omega_1}{2}} \times (1+z)^{3(\omega_0 + \omega_1)}~~~~~~~~.
\end{equation}
We find the observational data supported values of these parameters. To constrain the parameters of our universe we will take the help of the references \cite{Planck_2013_results_I, Planck_2013_results_XVI, Hinshaw}. Planck observation analysis takes the sum of neutrino masses fixed to 0.06eV, while the Wilkinson Microwave Anisotropy Probe (WMAP) sets it to zero. The perturbation amplitude $\bigtriangleup_R^2$ is specified at the scale $0.05 ~Mpc^{-1}$ for Planck data but $0.002~ Mpc^{-1}$ for WMAP, so the spectral index $n_s$ needs to be taken into account in comparing them. In the reference \cite{Lahav_Liddle}, uncertainties are shown at $68\%$ confidence.

Barboza-Alcaniz parametrization does not belong to the so called two redshift parametrization families, i.e., $\omega(z) = \omega_0 + \omega_1 \left(\frac{z}{1+z} \right)^n$ and $\omega(z) = \omega_0 + \omega_1 \frac{z}{(1+z)^n}$. It only generalizes to the linear parametrization for high redshift. Else it has natures completely independent from that the above two families and particular members of them (viz CPL, JBP, etc.). Our motivation is to highlight different cosmological evolutionary properties shown by this particular Barboza-Alcaniz EoS with a continuous data set. Besides Einstein's general relativity we wish to study this EoS in the background of Loop Quantum Gravity and Horava Lifshitz Gravity as this quantum gravity theories which do not possess any future singularities and this is why the Barboza-Alcaniz redshift prametrization may show some new results in the background of such four dimensional quantum gravity theories. We wish to check which gravity constrains this redshift parametrization parameters most. A comparative study of the behaviours of $\omega_0$ and $\omega_1$ in the general relativity along with different modified gravities will be done.

In this letter, at first we will discuss about Einstein's Gravity and will plot $1\sigma$, $2\sigma$ and $3\sigma$ confidence contours in $\omega_0-\omega_1$ plane for the best fit values of $\{H(z)-z\}$, $\{H(z)-z\}$ + BAO and $\{H(z)-z\}$ + BAO + CMB data respectively. Then we will discuss the same for LQG and HL Gravity. Finally, we will discuss in brief the results which we have got and represent a comparative study and conclude.

\section{Einstein's General Relativity : Constraints on Barboza Alcaniz parameters}
 General relativity generalizes special relativity and Newton's law of universal gravitation. In this theory, the curvature of space-time is directly related to the quantity of energy and momentum of the matter and radiation present in the concerned space-time. The relation is governed by the Einstein's field equations. GR possesses some local singularities (like black holes etc) and some past or future cosmological singularities like ``Big Rip", ``Big Bang" etc. Newtonian mechanics is good in small area, whereas Einstein's general relativity is very much required while we are to analyze the universe as a whole. We can find several articles where different DE EoS parameters are constrained in the background of general relativity \cite{Linear, CPL, JBP, Log, ASSS1, ASSS2}.

We will consider FLRW universe and Einstein's field equations turn to be
\begin{equation}\label{Ads_field_equation_I}
\bigg(\frac{\dot{a}}{a}\bigg)^2 + \frac{kc^2}{a^2} = \frac{8 \pi G}{3} \rho_{tot}
\end{equation}
and
\begin{equation}\label{Ads_field_equation_II}
\frac{2\ddot{a}}{a} + \bigg(\frac{\dot{a}}{a}\bigg)^2 + \frac{k c^2}{a^2} = -\frac{8 \pi G}{c^2} p_{tot}~~~~~.
\end{equation}
From (\ref{Ads_field_equation_I}) we find ,
$$H^2 = \bigg(\frac{\dot{a}}{a}\bigg)^2 = \frac{8 \pi G}{3}(\rho_{rad}  + \rho_{DM} + \rho_{DE}) - k c^2 (1+z)^2$$
\begin{equation}\label{Ads_field_equation_III}
= H_0^2 \bigg[ \Omega_{rad}(1 + z)^4 + \Omega_{DM}(1 + z)^3 + \Omega_{DE}(1 + z)^{3(\omega_0 + \omega_1)}\bigg\{\frac{1 + z^2}{(1 + z)^2}\bigg\}^{\frac{3\omega_1}{2}} \bigg]
\end{equation}
where $\Omega_i = \frac{8 \pi G}{3 H_0^2} \rho_i$, $i = rad, DM$ and $DE$ are the dimensionless density parameters.

Now, at $z = 0$, we have $H = H_0$ and for flat space we get from (\ref{Ads_field_equation_III}) the form of $\Omega_{DE}$ as \cite{Spergel}
\begin{equation}
\Omega_{DE} = 1 - \Omega_{rad} - \Omega_{DM} + \frac{k c^2}{H_0^2} = 0.0439688
\end{equation}

In this part of letter, we will like to very much to tell about the evolution of the value of value Hubble's constant, rather Hubble's parameter while it was measured again
and again via different data, tool and methodologies. Measurement of $H_0$ $(=72 \pm 8)$ based upon Sunyaev-Zel'dovich effect was done by Hubble Space Telescope key Project \cite{Planck_2013_results_XVI} in 2001-05. In 2006-08, Chandra X-ray Observatory, using the same methodology of Sunyaev-Zel'dovich effect speculated $H_0$ as $76.9^{+10.7}_{-8.7}$ \cite{Spergel1}. Upto 2009, WMAP (5 years) only way determined $H_0$ to be $71.9^{+2.6}_{-2.7}.$ However in 20/12/2012 the same was determined to be equal to $69.32 \pm 0.80$ by WMAP (9 years) \cite{Boyle}. Plank Mission, on 21/03/2013 has establish $H_0 = 67.80 \pm 0.77$ \cite{Tully}. At the time of drafting this letter, we have taken Hubble space telescope and Gaia's determination of $H_0$ as $73.52 \pm 1.62$ \cite{Planck_2018, Riess1}. But through the period ofreview process we are acquainted with Dark Energy Survey (DES) collaboration data (visible and near infrared story using 4 meter Víctor M. Blanco Telescope, Chile) which uses supernova measurements using the issue distance ladder method based on Baryon acoustic oscillations and determines $H_0$ as $67.77 \pm 1.30$ \cite{Chen_H_Y} in 06/11/2018.
\begin{equation}\label{value_of_dimensionless_parameters_and_Hubble_constant}
\left. \begin{aligned}
Cold~ dark~ matter~ density~ \cite{Palash}~ : \Omega_{CDM}~ = (0.112 \pm 0.006) \times (0.704 \pm 0.025)^{-2} \\
Radiation~ density~ \cite{Lahav_Liddle}~ : \Omega_{rad}~ = (2.47 \times 10^{-5}) \times (0.704 \pm 0.025)^{-2} \\
Hubble's~ constant~ \cite{Macaulay}~ : H_0 = 67.77~ \pm~ 1.30~ km/s/Mpc
\end{aligned}\right\rbrace~~~~~~.
\end{equation}

{\bf Different Confidence Contours In $\omega_0-\omega_1$ Plane : Einstein's Gravity Is Concerned}

We will tabulate the best fit values for $\omega_0$, $\omega_1$ and corresponding $\chi^2$ using $H(z)-z$ data, $H(z)-z$ data+BAO and $H(z)-z$ data+BAO+CMB respectively in Table-II. To support the observational data from expanding universe whenever $\omega_0$ is negative or very particularly near to $\omega_1$, we speculate that our model is terminally supporting $\Lambda$CDM model. But here we observe $\omega_1$ is preferable to have negative values when we include high redshift data, positivity of $\omega_1$, as it is attached with $z$ and $z^2$ terms, may not lead us to a negative pressure from the EoS $\omega = \frac{p}{\rho}$. This is leading us to negative valued $\omega_1$ as a best fit.

~~~~~~~~~~~~~~~~~~~~~~~~~~~~~~~~~~~~~~~~~~~~~~~~~~~~~~~~~~~~Table-II~~~~~~~~~~~~~~~~~~~~~~~~~~~~~~~~~~~~~~~~~
\begin{center}

\begin{tabular}{|| c | c | c | c ||}
\hline
Tools &  $\omega_0$ & $\omega_1$ & $\chi^2$ \\ [0.5ex]
\hline
$H(z)-z$ data & -1.01168 & 2.73165 & 40.6989 \\
\hline
$H(z)-z$ data + BAO &  -0.994854 & 2.42484 & 801.942 \\
\hline
$H(z)-z$ data + BAO + CMB & -0.992483 & 2.5234 & 9995.99 \\
\hline 
\end{tabular}
\end{center}
We have plotted the $1\sigma, 2\sigma,$ and $3\sigma$ confidence contours in $\omega_0 - \omega_1$ plane. We have done it for $\{H(z) - z\}$ data (fig 1a), $\{H(z) - z\}$ data + BAO (fig 1b) and $H(z) - z$ data + BAO + CMB (fig 1c). The general natures of the contours for all these three cases  are same. We see it to be highly eccentric oval shaped. If $\omega_0$ is increased and $\omega_1$ is decreased, it is seen that we can vary the domain for a large range to stay within the $1\sigma$ confidence. However, this domain of $1\sigma$ is approximately $-4 < \omega_0 < 3.5, ~~-45 < \omega_1 < 22$. If BAO is included, the range towards the high $\omega_0$ and low $\omega_1$ is increased a bit [upto $--4.15 < \omega_0 < 3.7$, $-46 < \omega_1 < 22$]. This trend stays on for inclusion of CMB as well [upto $-4.1 < \omega_0 < 3.6$, $-46 < \omega_1 < 22$]. However, we can conclude that a huge range of $\omega_0$ and $\omega_1$ supports the $1\sigma$ confidence for this particular type of parametrization.

\begin{figure}[h!]
\begin{center}
~~~~~~Fig.-1(a) ~~~~~~~~~~~~~~~~~~~~~~~~~~~~~~~Fig.-1(b)~~~~~~~~~~~~~~~~~~~~~~~~~~~~~~~Fig.-1(c)~~~~~~\\
\includegraphics[scale=.59]{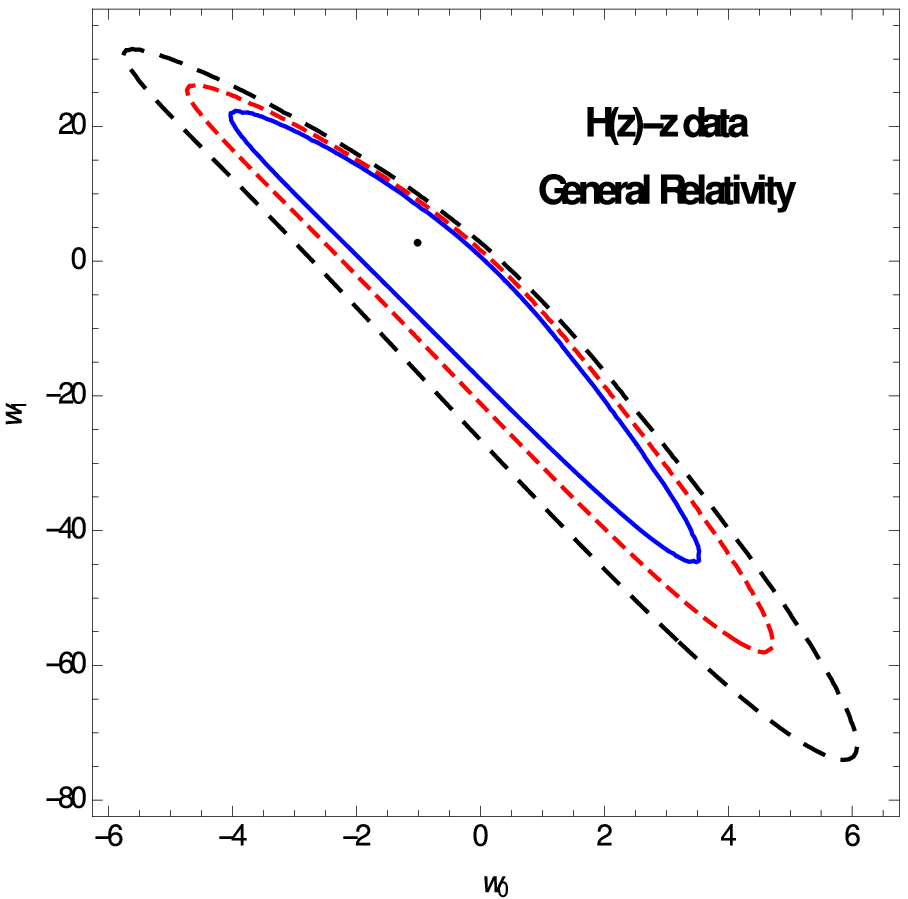}~~
\includegraphics[scale=.59]{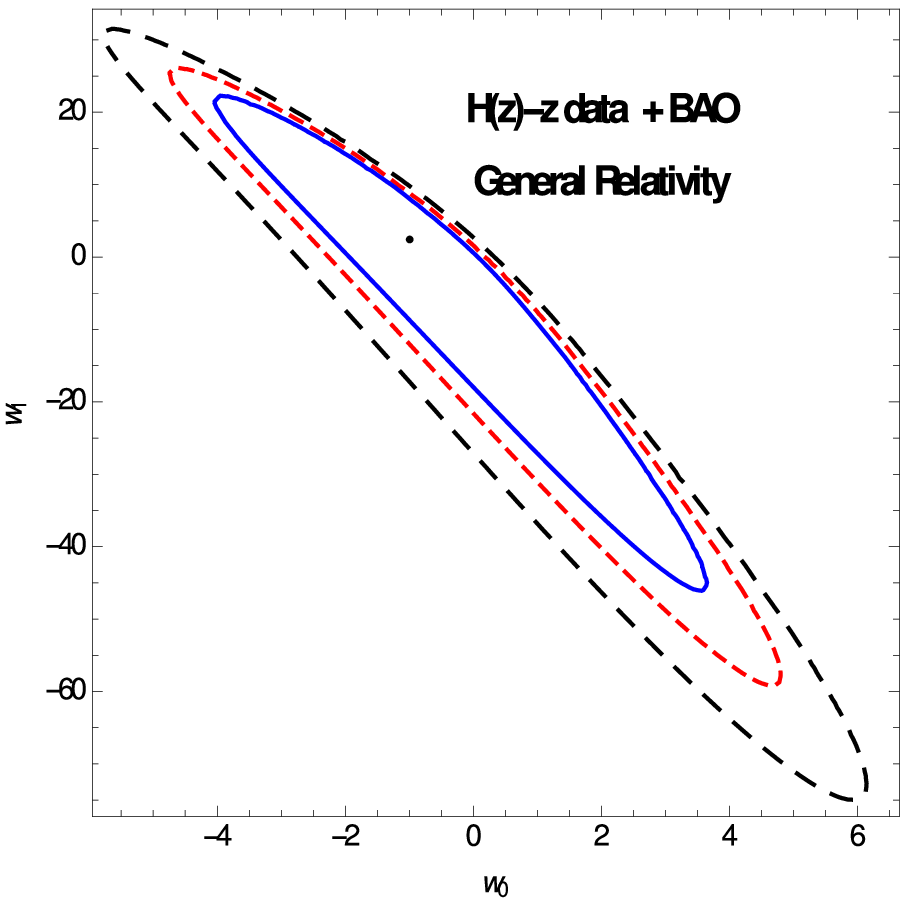}~~
\includegraphics[scale=.59]{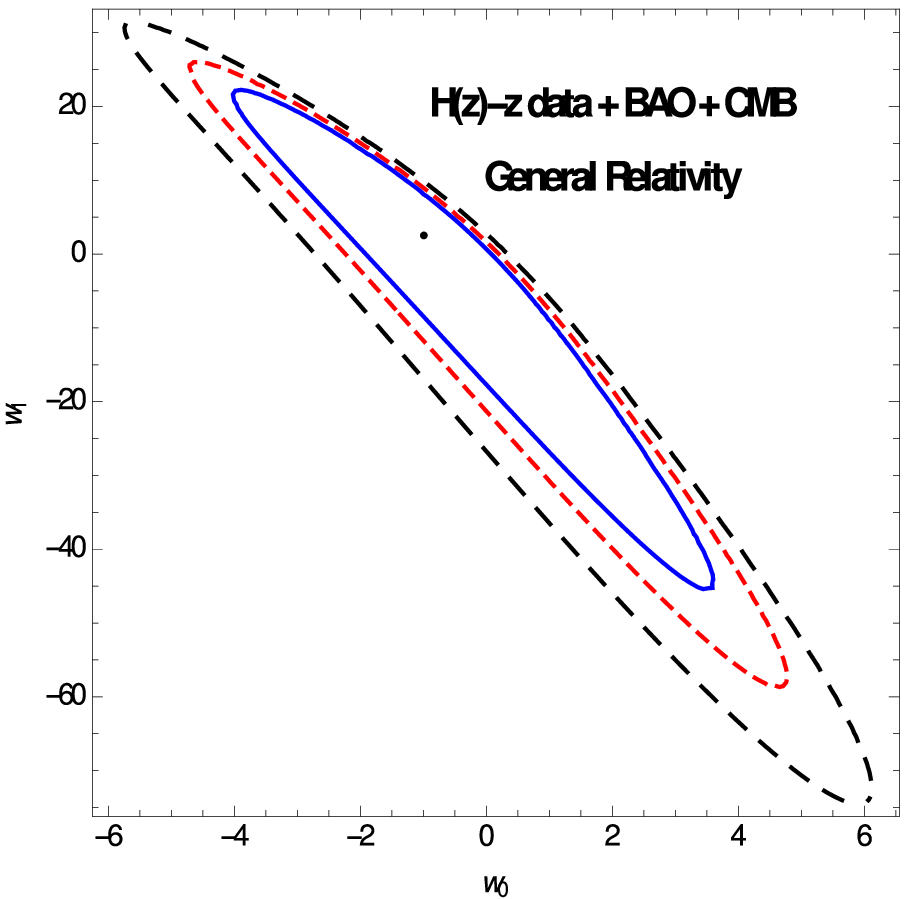}~~\\
Fig-$ 1(a)$-$(c) $ represent the $1\sigma$, $2\sigma$ and $3\sigma$ confidence contours in $\omega_0-\omega_1$ space for simply $H(z)-z$ data set, $H(z)-z$ data set+BAO and $H(z)-z$ data set+BAO+CMB respectively : FLRW metric in Einstein's General Relativity is considered. $\Omega_{CDM}$, $\Omega_{rad}$, $H_0$ are taken as given in equation (\ref{value_of_dimensionless_parameters_and_Hubble_constant}).\\
\end{center}
\end{figure}

It is noted that if we fix any one of $\omega_0$ or $\omega_1$, we can not increase the rest and stay inside the $1\sigma$ confidence. But increase of $\omega_0$ is supported along with decrease of $\omega_1$ or the reverse. As we involve BAO we are independent to move to both the extremities (i.e., low $\omega_0$ along with high $\omega_1$ or the opposite) for some more larger domain. Again, inclusion of BAO and CMB both allows us to move more towards low $\omega_0$ and high $\omega_1$, but we are not free to move more towards low $\omega_1$ and high $\omega_0$ than the $\{H(z)-z\}+BAO$ case. So when we take only $H(z)-z$ data set, it forces us to stay in $\omega(z)< 0$ region unless we choose to be with the high $\omega_0$, low $\omega_1$ case for $z > 0.1$ (approximately). So present time universe is negative pressure dominated. Inclusion of BAO increases the domain of ($\omega_0,~\omega_1 $) in both sides. Interesting phenomena is observed only if BAO and CMB both are added when we see negative $\omega_0$ is more likely to occur. If we consider general relativity, we will get a negative pressure creating agent when Barboza-Alcaniz parametrization is considered.
\section{View from Loop Quantum Gravity}
The renowned cube of physics was constructed by considering classical mechanics at one of its vertex which is taken to be the origin. Vertex on X axis may be treated as the shift towards the special relativity. Vertices on Y and Z axes can be treated as the shifts towards the Quantum Mechanics and Newtonian Gravity respectively. Vertices on XZ plane and XY plane will denote general relativity and relativistic quantum theories respectively. Vertex diagonally opposite to the origin will denote the theory of everything which is yet to reach. There are many possible proposed candidates towards {\it ``theory of everything"}. Loop quantum gravity is one among these possible candidates of theory of everything where we try to merge quantum mechanics with general relativity. More clearly, in this theory, our goal is to unify gravity with other three fundamental forces of nature. As we stated in the previous section, Einstein's views towards gravity were to treat it not as a force created by so called Newton's gravitation but to sense it as a property of space-time (or more specifically ripples of space-time created by the matter energy present in the space-time) itself.

In LQG, we attempt to develop a quantum theory of gravity based directly on Einstein's geometric formulation. In this theory, temporal and spatial coordinates are quantised and these coordinates are treated granular and discrete because of the quantization (like photons in second quantization of electromagnetic waves). Most developed application of LQG is done in cosmology, popularly known as Loop Quantum Cosmology (LQC). LQC possesses the prospective of non perturbative and background independent quantisation of gravity \cite{LQG,Bojowald,Ashtekar1,Ashtekar2, Ashtekar3, Rovelli2}. Many exotic matter models have been studied in the scenario of LQC \cite{Wu,Chen,Jamil,Fu,Chakraborty,Biswas_R}. This theory has a robust utility in studying early universe and physics of ``Big Bang", evolution process of universe (inflation, deceleration phase, accelerating phase) and even future singularities like ``Big Rip", ``Big Crunches" etc. These singularities at semi classical regime can be avoided in LQC. Along with these features, the modification in standard FLRW cosmology due to LQC is more dominant and the universe starts to bounce and to oscillate forever.

Due to the extreme smallness of the Planck length, quantum gravity effects are difficult to measure. Recent gravitational wave detections \cite{Gravitational_Waves}, however, motivated physicists to consider the possibility of measuring quantum gravity effects. This is why we choose this model (which is a 4 Dimensional theory as well) as the back ground of our data analysis for exotic matter EoS parameters in the late time cosmic acceleration.

The first announcement of LQG was introduced in an international conference in India in 1987 \cite{Rovelli}. LQG is a mathematically well defined non-perturbative and background independent quantisation of GR with its conventional matter couplings. Some major benefits of LQG is that the theory has been proved finite in a more definitive sense and computation of the physical spectra of geometrical quantities such as area and volume which allows quantitative predictions on Planck-scale physics. Derivation of the Bekenstein-Hawking BH entropy formula was obtained from LQG model \cite{LQG}.

DE embedded in LQG is studied in many references like \cite{Wu, Chen, Jamil, Fu}. In this section we will constrain the DE parameters in the back ground of LQG. Considering the flat homogeneous and isotropic described by FLRW metric, the modified Einstein's equations in LQC read as
\begin{equation}\label{field_equation_with_critical_quantum_cosmology}
H^2 = \bigg(\frac{\dot{a}}{a}\bigg)^2 = \frac{\rho}{3}\bigg(1 - \frac{\rho}{\rho_c}\bigg)
\end{equation}
and
\begin{equation}
\dot{H} = -\frac{1}{2} (p + \rho) \bigg(1 - \frac{2 \rho}{\rho_c}\bigg)~~~~~~,
\end{equation}
where $\rho_c$ is critical loop quantum density as $$\rho_c = \frac{\sqrt{3}}{16 \pi^2 \gamma^2 G^2 \hbar}~~~~~,$$ $\gamma$ is the dimensionless Barbero-Immirzi Parameter.

From (\ref{field_equation_with_critical_quantum_cosmology}) we have,
$$H^2 = H_0^2 \bigg[\Omega_{rad}(1+z)^4 + \Omega_{DM}(1+z)^3 + \Omega_{DE}(1+z)^{3(\omega_0 + \omega_1)}\bigg\{\frac{1+z^2}{(1+z)^2}\bigg\}^{\frac{3\omega_1}{2}}\bigg]$$
\begin{equation}
\times\bigg[1 - \frac{3H_0^2}{\rho_c^2}\bigg(\Omega_{rad}(1+z)^4 + \Omega_{DM}(1+z)^3 + \Omega_{DE}(1+z)^{3(\omega_0 + \omega_1)}\bigg\{\frac{1+z^2}{(1+z)^2}\bigg\}^{\frac{3\omega_1}{2}}\bigg]~~~~~~~~.
\end{equation}
For terminal case at $z = 0$ we have,
\begin{equation}
1 = \bigg\{\Omega_{rad} + \Omega_{DM} + \Omega_{DE}\bigg\} \bigg[1 - \frac{3H_0^2}{\rho_c}(\Omega_{rad} + \Omega_{DM} + \Omega_{DE})\bigg]
\end{equation}

{\bf Confidence Contours In $\omega_0-\omega_1$ Plane : Loop Quantum Cosmology}


We will form Table-III with the best values of $\omega_0$, $\omega_1$ and $\chi^2$ using $\{H(z)-z$ data\}, $\{H(z)-z$ data\}+BAO and $\{H(z)-z$ data\}+BAO+CMB respectively.

~~~~~~~~~~~~~~~~~~~~~~~~~~~~~~~~~~~~~~~~~~~~~~~~~~~~~~Table-III~~~~~~~~~~~~~~~~~~~~~~~~~~~~~~~~~~~~~~~~~
\begin{center}
\begin{tabular}{|| c | c | c | c ||}
\hline
Tools &  $\omega_0$ & $\omega_1$ & $\chi^2$ \\ [0.5ex]
\hline
$H(z)-z$ data & -1.00492 & -5.05421 & 315.188 \\
\hline
$H(z)-z$ data + BAO & -1.00431 & -4.91369 & 1073.39  \\
\hline
$H(z)-z$ data + BAO + CMB & -1.00324 & -4.96937 & 10269.2 \\
\hline
\end{tabular}
\end{center}

We plot the $1\sigma$, $2\sigma$ and $3\sigma$ confidence contours in $\omega_0-\omega_1$ plane for $H(z)-z$ data (fig 2(a)), $H(z)-z$ data + BAO (fig 2(b)), $H(z)-z$ data + BAO + CMB (fig 2(c)), while LQC is accounted. Though the contours are a bit of oval shaped, the eccentricity is clearly less than GR plots. For inclusion of BAO, we see the contours are stretched in both ends (i.e., more $\omega_1$ with less $\omega_0$ and more $\omega_0$ with less $\omega_1$ ends). Inclusion of BAO and CMB stretches the contour more.

\begin{figure}[h!]
\begin{center}
~~~~~~Fig.-2(a) ~~~~~~~~~~~~~~~~~~~~~~~~~~~~~~~Fig.-2(b)~~~~~~~~~~~~~~~~~~~~~~~~~~~~~~~Fig.-2(c)~~~~~~\\
\includegraphics[scale=.59]{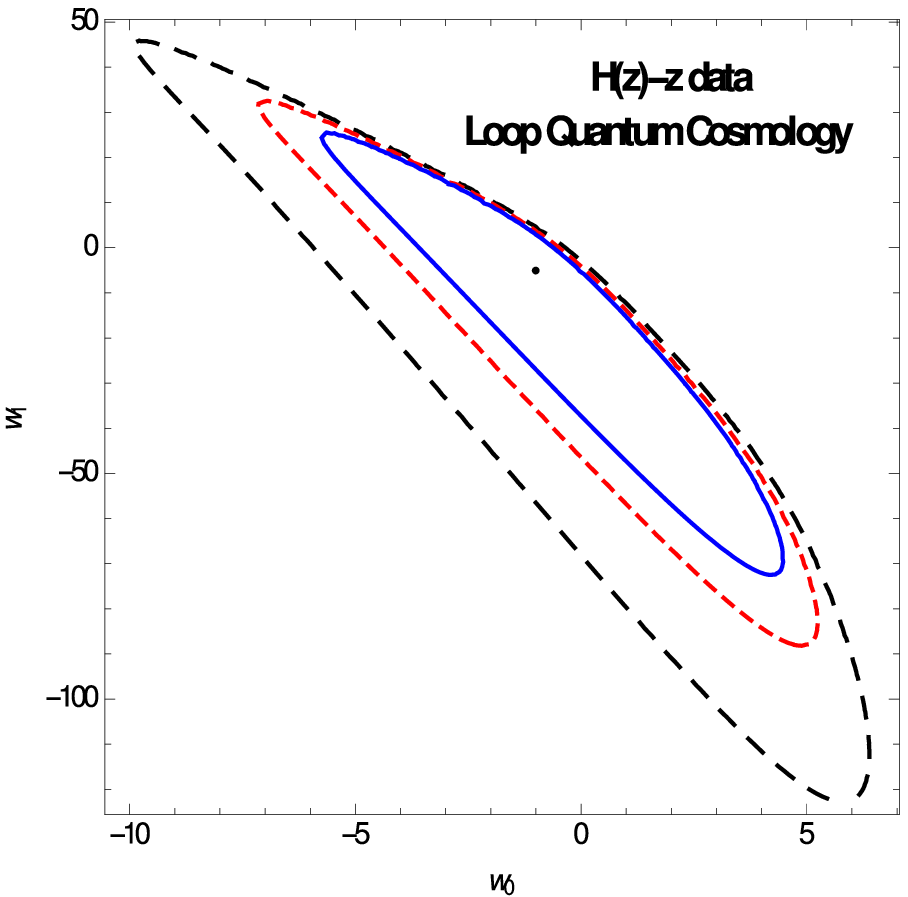}~~
\includegraphics[scale=.59]{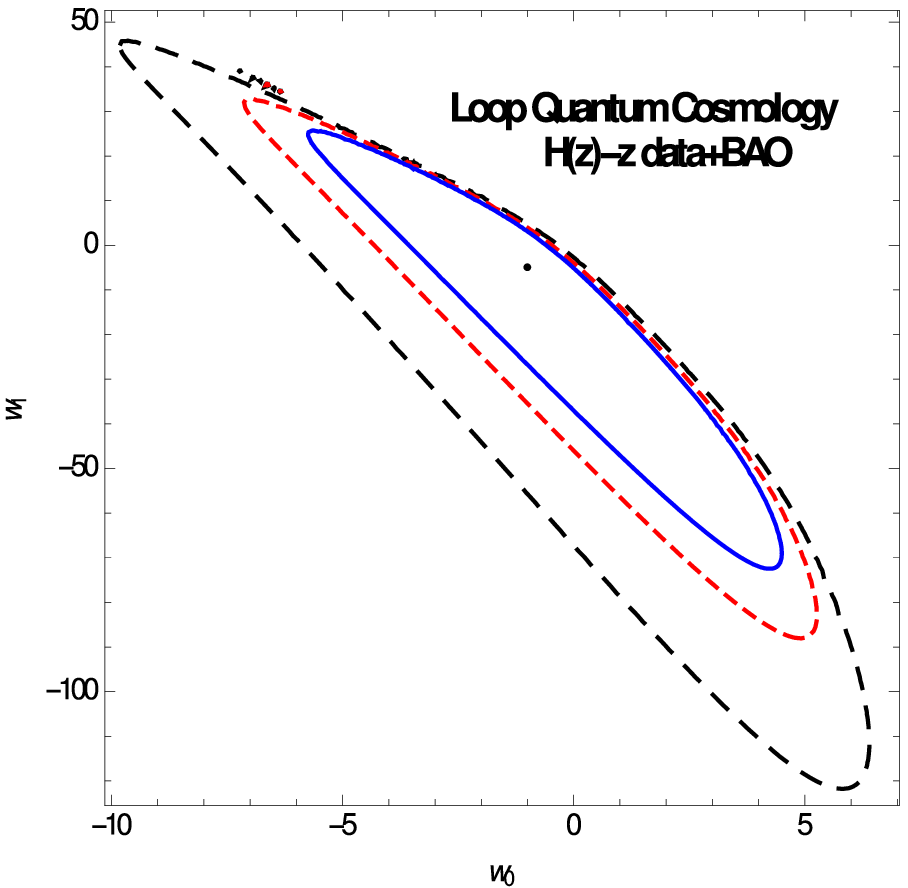}~~
\includegraphics[scale=.59]{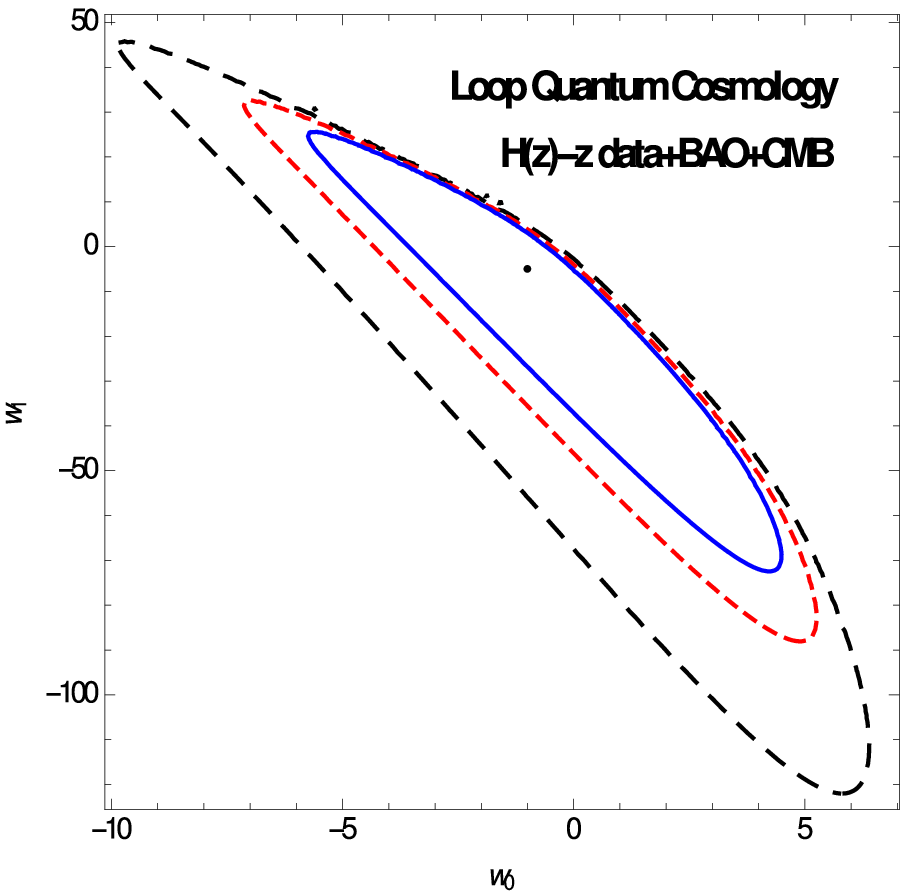}~~\\
Fig-$ 2(a)$-$(c) $ represent the $1\sigma$, $2\sigma$ and $3\sigma$ confidence contours in $\omega_0-\omega_1$ space for simply $H(z)-z$ data set, $H(z)-z$ data set+BAO and $H(z)-z$ data set+BAO+CMB respectively : FLRW metric in Loop Quantum Cosmology is considered. $\Omega_{CDM}$, $\Omega_{rad}$, $H_0$ are taken as given in equation (\ref{value_of_dimensionless_parameters_and_Hubble_constant}).\\
\end{center}
\end{figure}

First thing to be pointed in LQC is that the best fit point is in the third quardrant, i.e., for every positive $z$ we will have a negative EoS $\omega(z)< 0$. However, while we wish to study the natures of the confidence contours, we observe that keeping $\omega_0$ constant at best fit, increment in $\omega_1$ is not allowed to a large extent. But we can decrease the value of $\omega_1$ keeping $\omega_0$ same. Again high $\omega_1$, low $\omega_1$ corner is narrower than the low $\omega_1$, high $\omega_0$ end. So the negativity in the EoS is merely increased by the $\omega_1 \frac{z(1+z)}{(1+z^2)}$ part and not at all dependent on the $\omega_0$ part only. Again if we compare this case with the general relativistic one, we observe that $1\sigma$ confidence domain of $\omega_0$ and $\omega_1$ for LQC is wider than GR. Cosmologically we can speculate from this tendencies in GR and LQC that to support the same data we can find LQC to more liberal on the changes of values of redshift parametrization parameters of Barboza-Alcaniz model. Inclusion of BAO along with the $H(z)-z$ data increases the domain to the low $\omega_1$ high $\omega_0$ corner. So in LQC inclusion of BAO put its effect on the coefficient of the $z$ associated terms. If we follow the best fit, at $z= 0$, i.e., at present time the EoS will be more negative by consideration of BAO. For inclusion of BAO and CMB both, the extreme ends go for but high $\omega_1$, low $\omega_0$ end is expanded more like general relativistic case. Apparently, it seems that the graphs are not changed even if the data sets are changed. But actually the $\chi^2$ values are changed a lot for different data. Similarly the range of $\omega_0$ and $\omega_1$ to stay within the $1\sigma$, $2\sigma$ and $3\sigma$ confidence changes due to the inclusion of BAO and CMB. Well, this is true that the overall pattern remains same. But if we see different redshift parametrization's studies in literature we can observe this tendency of having same pattern is followed \cite{Biswas_R,Biswas_R2,Biswas_R3}.
\section{With Horava Lifshitz Gravity}
While we are to construct the inflation epoch, i.e., are closed to the Planck era, a Ultraviolet (UV) complete theory is required to be built. Horava- Lifshitz  (HL) gravity \cite{Horava2, Horava1, Lifshitz} is a milestone in this particular field.

Lorentz invariance may not exist quantum mechanically due to its nature of continuous symmetry of space-time. It is quiet reasonable that Lorentz invariance is broken in the UV cut off but recover later in the infra red (IR) cut off. While Lorentz invariance is broken, higher order spatial order derivative operators can be included into the corresponding Lagrangian to improve the UV cut off behaviour. Besides the time derivative operators are to be kept to the 2nd order in order to evade this pointed out by Ostrogradsky’s ghosts. This is the methodology followed by Horava \cite{Horava2}.

Horava has chosen to break the Lorentz invariance by considering anysotropy scaling between time and space given as $t \longrightarrow b^{-\mathcal{Z}}t, x^i \longrightarrow b^{-1}x'^i$, $(i = 1,2,...,d)$ where $\mathcal{Z}$ is the dynamical critical exponent. $d$ is the spatial dimension of the space-time \cite{Horava2,Visser1,Visser2,Anselmi,Fujimori1,Fujimori2}. For Lorentz invariance, $\mathcal{Z}= 1$, while $\mathcal{Z}\geq d$ is required for power-cutting renormalizability (The total effective mass of a spherical charged particle includes the actual bare mass of the spherical shell (in addition to the mass mentioned above associated with its electric field). If the shell's bare mass is allowed to be negative, it might be possible to take a consistent point limit. This is called `Renormalisation'). Generally, we consider $d= 3$ and take the minimum value $\mathcal{Z}= d$, except in particular consideration.

Horava gravity may be constructed with projectability condition where the Hamiltonian constraint becomes global from which it may mimic dark matter emerging as an integrating constant of dynamical equation. A nontrivial generalisation may be embedded into string theory by using non relativistic AdS/ CFT correspondence \cite{Janiszewski1, Janiszewski2}. This leads to non-projectability counter-part of Horava gravity.

Gravitational collapse of a spherically symmetric object was studied in \cite{Greenwald}. Different collapsed objects have been pointed out \cite{Goldoni1, Goldoni2, Goldoni3}.

Due to Lorentz violation, the most oppressive constraint has considered in the preferred frame as it is noticed by Blas, Pujolas and Sibiryakov (BPS hereafter) who first  introduced in the non projectable case \cite{BPS} which require \cite{BPS2, Will}
\begin{equation}
|\lambda - 1| \leq 4 \times 10^{-7}, M_* \lesssim 10^{15}~GeV~~,
\end{equation} 
where $\lambda$ is coupling constant and $M_*$ is a new energy scale. To obtain the previous constraints, BPS used the results of Einstein-aether theory, as these theories coincide in the IR \cite{Jacobson, Jacobson2}.

Though the most oppressive constraints of the theory were obtained \cite{Yagi, Yagi2}, the limits from binary pulsars had been also studied recently. When the solar system tests is saturated, the allowed range of the preferred frame effects the limit for $\lambda$ is given equation from (16), so the maximum bound $M_*$ remains the same.

Even it is found from different observation of GW from the events GW150914 and GW151226 that HL gravity is compatible enough and moderate constraints of its different parameters were obtained. Still there are many un-answered questions regarding these gravity theory. However this seems to be a promising alternative in quantisation in HL gravity.

We obtain the Friedmann equations as \cite{Jamil_HL, Paul}
\begin{equation}
\frac{H^2}{\kappa^2} = \frac{1}{6(3\lambda -1)}(\rho_m + \rho_r) + \frac{1}{6(3\lambda -1)}
\Bigg[ \frac{3\kappa^2 \mu^2 k^2}{8 (3 \lambda -1)a^4} + \frac{3 \kappa^2 \mu^2 \Lambda^2}{8 (3\lambda -1)}\Bigg] - \frac{\kappa^2 \mu^2 \Lambda k}{8 (3 \lambda -1)^2 a^2}~~~~~~~~~~~~~~~~~~and
\end{equation}
\begin{equation}
\dot{H} + \frac{3H^2}{2} = -\frac{\kappa^2}{4(3\lambda -1)}(\rho_m \omega_m + \rho_r \omega_r) - \frac{\kappa^2}{4(3\lambda -1)} \Bigg[\frac{3 \kappa^2 \mu^2 k^2}{8(3\lambda -1)a^4} + \frac{3 \kappa^2 \mu^2 \Lambda^2}{8(3\lambda -1)}\Bigg] - \frac{\kappa^4 \mu^2 \Lambda k}{8(3\lambda -1)^2 a^2}~~~~~~~~.
\end{equation}
From detailed balance, the first Friedmann equation can be written as
$$H^2 = \frac{8 \pi G}{3}\rho_{tot} + \bigg( \frac{k^2}{2 \Lambda a^4} + \frac{\Lambda}{2}\bigg) - \frac{k}{a^2}~~~~~~~.$$
Writing in detail we have,
\begin{equation}
H^2 = H_0^2 \bigg[ \Omega_{rad}(1+z)^4 + \Omega_{DM}(1+z)^3 + \Omega_{DE}(1+z)^{3(\omega_0+\omega_1)}\bigg\{\frac{1+z^2}{(1+z)^2}\bigg\}^{\frac{3\omega_1}{2}} + \bigg\{\Omega_{\Lambda} + \frac{\Omega_k^2(1+z)^4}{4\Omega_{\Lambda}}\bigg\} + \Omega_k(1+z)^2 \bigg]~~~~~~,
\end{equation}
where $\Omega_i \equiv \frac{8 \pi G}{3 H_0^2}\rho_i$, $\Omega_k = -\frac{k}{H_0^2 a^2}$, $\Omega_{\Lambda} = \frac{\Lambda}{2 H_0^2}$
and for $z = 0$, we have,
\begin{equation}
1 = \Omega_{rad,0} + \Omega_{DM,0} + \Omega_{DE,0} + \Omega_{\Lambda} + \frac{\Omega_k^2}{4 \Omega_{\Lambda}} + \Omega_k
\end{equation}

{\bf Confidence Contours In $\omega_0-\omega_1$ Plane : Horava Lifshitz Gravity}


Now we will construct Table-IV with the best fit values of $\omega_0$, $\omega_1$ and $\chi^2$ using $\{H(z)-z$ data\}, $\{H(z)-z$ data\}+BAO and $\{H(z)-z$ data\}+BAO+CMB respectively.
 
~~~~~~~~~~~~~~~~~~~~~~~~~~~~~~~~~~~~~~~~~~~~~~~~~~~~~~Table-IV~~~~~~~~~~~~~~~~~~~~~~~~~~~~~~~~~~~~~~~~~
\begin{center}
\begin{tabular}{|| c | c | c | c ||}
\hline
Tools &  $\omega_0$ & $\omega_1$ & $\chi^2$ \\ [0.5ex]
\hline
$H(z)-z$ data & -1.01043 & 3.0142 & 38.0163\\
\hline
$H(z)-z$ data + BAO & -1.00456 & 2.7304 & 799.849  \\
\hline
$H(z)-z$ data + BAO + CMB & -1.00427 & 2.94723 & 9993.65  \\
\hline
\end{tabular}
\end{center}

We plot the $1\sigma$, $2\sigma$ and $3\sigma$ confidence contours in $\omega_0 - \omega_1$ plane for $H(z)-z$ data (fig 3a), $H(z)-z$ data + BAO (fig 3b), $H(z)-z$ data + BAO + CMB (fig 3c) in the back ground of Horava Lifshitz gravity. We see the general trend is eccentric oval. But the contours for HL gravity are less eccentric than general relativity and more eccentric than LQC.
\begin{figure}[h!]
\begin{center}
~~~~~~Fig.-3(a) ~~~~~~~~~~~~~~~~~~~~~~~~~~~~~~~Fig.-3(b)~~~~~~~~~~~~~~~~~~~~~~~~~~~~~~~Fig.-3(c)~~~~~~\\
\includegraphics[scale=.59]{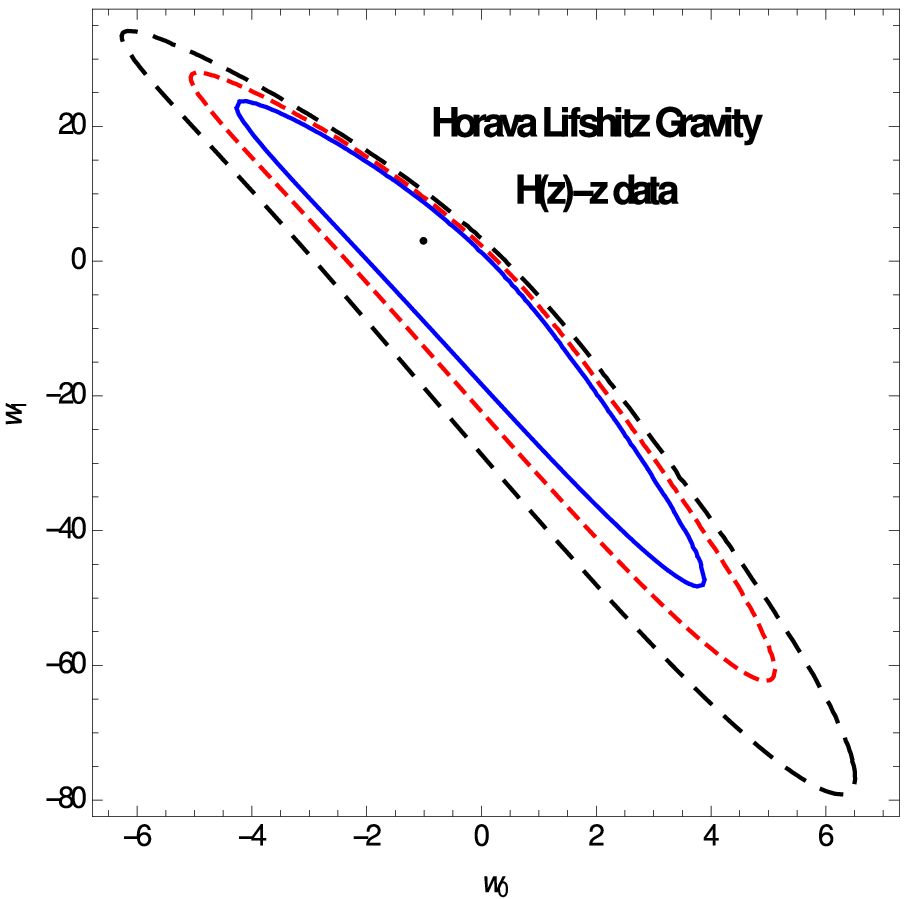}~~
\includegraphics[scale=.59]{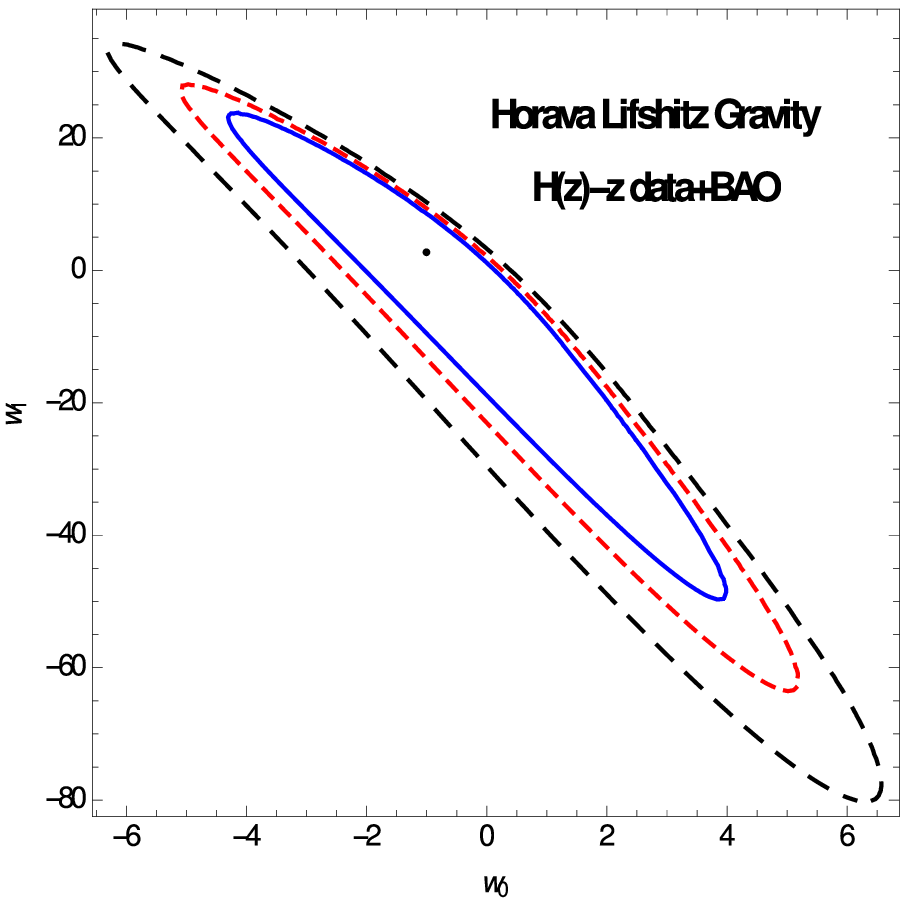}~~
\includegraphics[scale=.59]{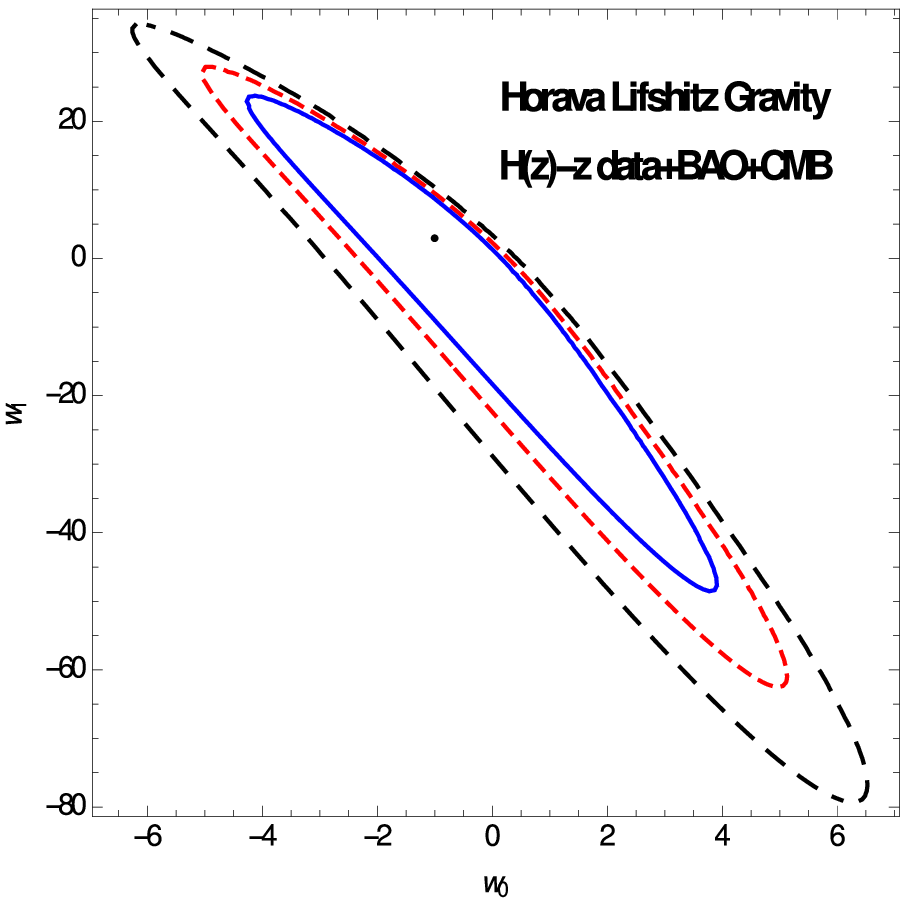}~~\\
Fig-$ 3(a)$-$(c) $ represent the $1\sigma$, $2\sigma$ and $3\sigma$ confidence contours in $\omega_0-\omega_1$ space for simply $H(z)-z$ data set, $H(z)-z$ data set+BAO and $H(z)-z$ data set+BAO+CMB respectively : FLRW metric in Horava Lifshitz gravity is considered. $\Omega_{CDM}$, $\Omega_{rad}$, $H_0$ are taken as given in equation (\ref{value_of_dimensionless_parameters_and_Hubble_constant}).\\
\end{center}
\end{figure}

For Horava Lifshitz gravity the confidence contours are narrower than those of LQC, but wider than those of GR. This gravity theory belongs somewhere in between GR and LQC, while it is a question to constrain Barboza-Alcaniz parameters under the $H(z)-z$ data tabulated in Table-I. Inclusion of BAO stretches the extreme ends of confidence contours like other gravity theories. Even the inclusion of CMB and BAO also shows the same pattern as before.
\section{Redshift-Magnitude Observations from SNeIa Data :}
Since 1995, two different collaborating teams of high redshift supernova search and supernova cosmology project have started to discover several types highly redshifted distant type Ia supernovae \cite{cosmic_acceleration_paper_1,Perlmutter2}. These observations were able to measure the distance modulus of a supernova of its redshift $z~$\cite{Riess2,Kowalaski}. In this section, we have considered 557 different SneIa observations data (Union2 sample \cite{Amanullah}). The DE density determined from the luminosity distance $d_{L}(z)$ (from the observations) helps us to construct the formula distance as
\begin{equation}
d_{L}(z) = (1+z)H_0 \int_0^z \frac{dz'}{H(z')}~~~~~~~~~~~~.
\end{equation} 
Again the apparent magnitude $m$ and the redshift $z$ of a supernova can be directly measured from observations. The apparent magnitude $\mu$ is related to the luminosity distance $d_L$ of a supernova by the relation
\begin{equation}\label{Mu(z)}
\mu(z) = 5log_{10}\left[\frac{d_L(z)/H_0}{1 MPc}\right]+25~~~~~~~~~~~.
\end{equation}
For our theoretical model and Union2 sample of SneIa supernova we plot the best fit of distance modulus as a function $\mu(z)$ of corresponding redshift in figures 4(a), 4(b), 4(c) for Einstein's General Relativity, Loop Quantum Cosmology and Horava- Lifshitz Gravity respectively.
\begin{figure}[h!]
\begin{center}
~~~~~~Fig.-4a ~~~~~~~~~~~~~~~~~~~~~~~~~~~~~~~Fig.-4b~~~~~~~~~~~~~~~~~~~~~~~~~~~~~~~Fig.-4c~~~~~~\\
\includegraphics[scale=.63]{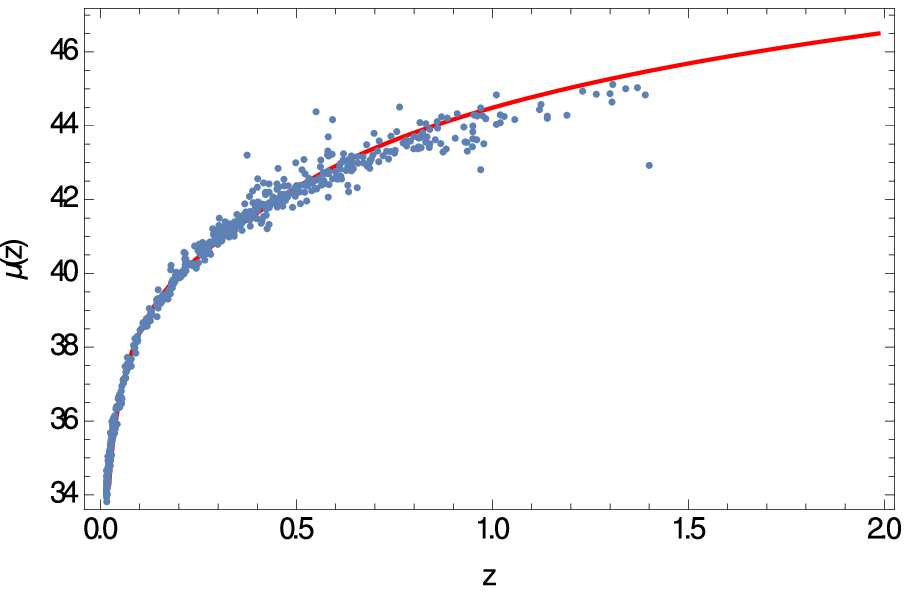}~~
\includegraphics[scale=.63]{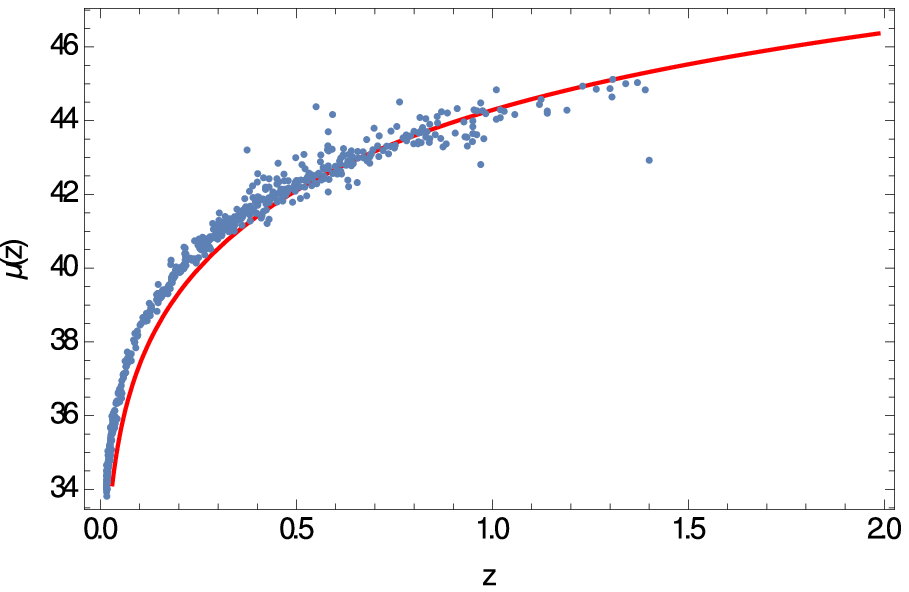}~~
\includegraphics[scale=.63]{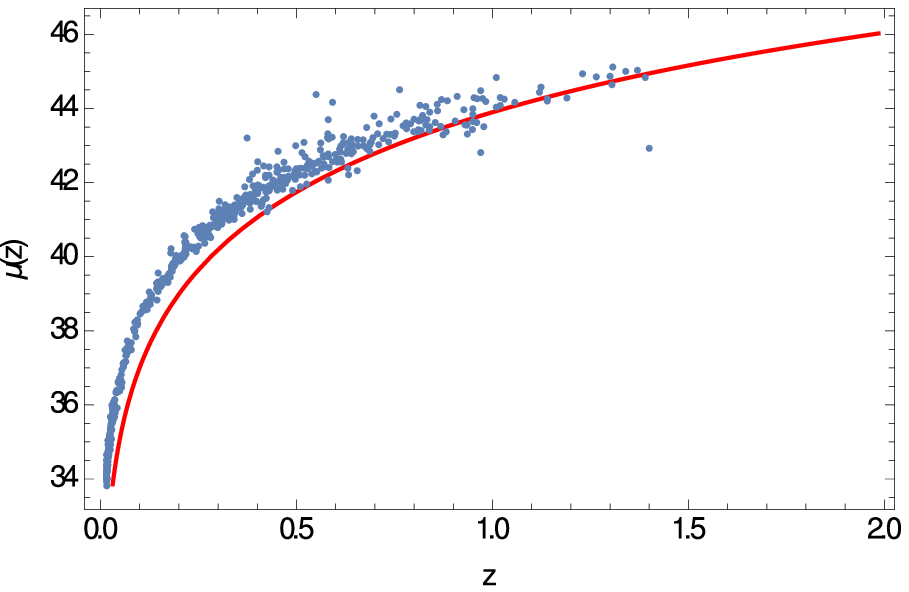}~~\\
Fig-$ 4(a)$-$(c)$ represent the variation of $\mu(z)$ defined by the equation (\ref{Mu(z)}) with respect to $z$ for Barboza-Alcaniz parametrization (solid red line) in Einstein's General Relativity, Loop Quantum Cosmology and Horava- Lifshitz Gravity respectively. The blue dots denote the data points of the Union2 sample. \\
\end{center}
\end{figure}
In fig 4(a), we have plotted our model for supernova data. It seems that if $z > 1.1$, our model is over estimated. For $z < 1.1$, our model perfectly matches with the supernova data. While matching this model with with SNeIa data we see our model of Barboza-Alcaniz parametrization (BA) along with LQC is perfect for $z>0.5$. For $z<0.5$ our model is under determined. Match with supernova data says that BA parametrization in HL gravity is undetermined for the region $z<0.5$. As a whole this model can not be taken a good fit to the supernova data.

\section{Brief Discussions and Conclusions :}
As we stated in our motivation, we wanted to study the Barboza-Alcaniz redshift parametrization of dark energy in different modified gravities by constraining the parameters under different $H(z)-z$ data along with the tools Baryonic Acoustic Oscillations and Cosmic Microwave Background. Firstly, we have done the constraining for Einstein's General Relativity. We have found the best fit values of parameters $\omega_0$ and $\omega_1$. We have plotted $68\% (1\sigma),~ 95.45\% (2\sigma)$ and $99.73\% (3\sigma)$ confidence contours in $\omega_0 - \omega_1$ plane.

It is interesting to note that the best fit points for different gravity theories are staying in different quadrants. In Einstein General relativity,  it is situated in $2^{nd}$ quadrant and addition BAO and CMB changes the range of $\omega_0$ and $\omega_1$. The  contour stretches keeping its core shape unchanged. From the end point of this uneven oval-disk types contour, we can note the different ranges of $\omega_0$ and $\omega_1$ in different cases. From the variation of ranges we can figure out the $1\sigma$ confidence level region in Einstein's gravity for Barboza-Alcaniz parametrization. So we can conclude for high $z$ the model may give positive $\omega(z)$ and may create positive pressure. However, our present time ($z= 0$) universe is in negative pressure dominated era. In LQC, we can see that the best fit points are always in $3^{rd}$ quadrant for all our derived cases. More noticeable incident is that the $1\sigma$ confidence contour does not stretch so much. It remains almost same after adding the tools BAO and CMB. Keeping the best fit values in $3^{rd}$ quadrant, it always remains negative and indicates to create negative pressure. But regarding comparison with Einstein's General relativity case, we should mention LQC's  $1\sigma$ confidence contour is wider than that of GR. The range of $\omega_0$ and $\omega_1$ is higher rather than that of GR. In HL gravity, the best fit value is situated in $2^{nd}$ quadrant. The $1\sigma$ confidence contours are less wider than LQC and almost same even if BAO and CMB are included. Studying three gravity theories for Barboza-Alcaniz parametrization we can speculate that the $1\sigma$ confidence contour in LQC is more wider than the rests and the best fit points stay in $3^{rd}$ quadrant and it never violate the negative pressure dominance property in the present time.  Actual beauty of Barboza-Alcaniz parametrization is that it does not shift $\omega_0$ and $\omega_1$ together towards high or low values, i.e., the parametric values are allowed to move in such a region that the ultimate pressure stays negative. So comparatively, we can see the GR and HL gravity show same kind of properties with this parametrization. LQC, however, is dark energy dominated and the extent of the dominance is more than other two gravity theories.

Theoretically, the present day true value of cosmological constant should be equal to $-1$. But generally it differs when attractive time-varying forms of vacuum energy viz quintessence etc are taken into account. The present time value of $\omega = \frac{p}{\rho}$ (i.e., $\omega_0$) has measured by Planck Collaboration (2018) \cite{Planck_2018} as $\omega = -1.028 \pm 0.032$. However, this value was consistent with $-1$, if we assume no evolution in $\omega$ is there over cosmic time. We can conclude that the best fits derived by us are quite compatible with this range of values permitted by Planck's Collaboration (2018).

\vspace{.1 in}

{\bf Acknowledgment:}
This research is supported by the project grant of Government of West Bengal, Department of Higher Education, Science and Technology and Biotechnology (File no:- $ST/P/S\&T/16G-19/2017$). PB thanks Department of Higher Education, Science \& Technology and Biotechnology, West Bengal for Swami Vivekananda Merit-Cum-Means Scholarship. RB thanks IUCAA, Pune for Visiting Associateship. RB dedicates this article to his PhD supervisor Prof. Subenoy Chakraborty, Department of Mathematics, Jadavpur University, Kolkata-32, India to tribute him on his $60^{th}$ birth year.
  
\end{document}